\documentclass[aps,prb,twocolumn,superscriptaddress]{revtex4-1}
\usepackage{graphicx}
\usepackage{picture}
\usepackage{xcolor}
\usepackage{color}
\usepackage{tikz}
\usepackage{ulem}
\usepackage{amsmath}
\usepackage{rotating}

\begin{document}
\title{Thermalization of a Disordered Interacting System under an Interaction Quench}

\author{Eric Dohner}
\affiliation{Department of Physics, University at Albany (SUNY), Albany, New York 12222, USA}
\author{Hanna Terletska}
\affiliation{Department of Physics and Astronomy, Middle Tennessee State University, Murfreesboro, TN 37132, USA}
\author{Herbert F Fotso}
\affiliation{Department of Physics, University at Buffalo SUNY, Buffalo, New York 14260, USA}

\begin{abstract}
Although most studies of strongly correlated systems away from equilibrium have focused on clean systems, it is well known that disorder may significantly modify observed properties in  various nontrivial ways.  The nonequilibrium interplay of interaction and disorder in these systems thus requires further investigation.
In the present paper, we use the recently developed nonequilibrium DMFT+CPA embedding scheme, that combines both the dynamical mean field theory (DMFT) and the coherent potential approximation (CPA) nonequilibrium extensions, to characterize the relaxation and the thermalization of a disordered interacting system described by the Anderson-Hubbard model under an interaction quench. The system, initially in equilibrium at a given temperature, has the interaction abruptly switched from zero to a finite value at a given time. To investigate the role of disorder, we use our effective medium approach to calculate, for different values of the final interaction and of the disorder strength, the distribution functions as the system evolves in time. This allows us to determine the effective temperature after the quench and to analyze the effects of disorder on the thermalization for various interaction strengths. We find that, for moderate interactions after the interaction quench, disorder can tune the final temperature of the system across a broad range of values with increased disorder strength leading to lower effective temperature. 

\end{abstract}

\maketitle

\section{Introduction}

\noindent The  dynamics of quantum systems away from equilibrium has been the subject of increased interest as a result of the recent experimental advances extending from quantum information processing platforms to time-resolved spectroscopies.  A salient question that has garnered a great deal of attention is that of how quantum systems thermalize (or not) when they are abruptly driven out of equilibrium. Beyond the  theoretical question of how thermalization arises in quantum systems that are supposed to be governed in their dynamics by unitary time evolution operators\cite{Deutsch_PRA1994, Srednicki_PRE1994, RigolEtAl_Nature2008}, these research questions have important experimental consequences. For instance, it is often typical in the analysis of pump-probe spectroscopy experiments to use a so-called ``hot" electrons model whereby electrons are driven by the pump pulse into an equilibrated state that is thermalized at a higher temperature than that of the initial system\cite{PerfettiEtAl_PRL2006, PerfettiEtAl_PRL2007}. This brings into focus the importance of the relevant relaxation scenarios and the associated timescales. Also, experiments simulating various lattice models in optical lattices are either intrinsically out of equilibrium or can be used to simulate, through their high degree of tunability, the dynamics of nonequilibrium quantum systems\cite{BlochDaliwerger_RMP2008, Greinerbloch_Nat2002, BlochNatPhys2005, BakrGreiner_Nat2009}. This further highlights the need for accurate modeling and benchmarking.

While numerous efforts have been dedicated to the investigation of the thermalization of correlated quantum systems away from equilibrium\cite{DMFT_noneq, FK_NonEq_DMFT08, DMFT_noneq_Aoki}, little has been done to explore the effect of disorder which we can anticipate, in some circumstances, to have significant impacts on the dynamics\cite{Nandkishore_Huse_AnnRevCondMatPhys2015, Kondov_DeMarco_PRL2015} and which we know to be ubiquitous in most systems of interest.
In particular, nonequilibrium dynamical mean field theory (DMFT) was used to investigate the thermalization of correlated systems in a variety of nonequilibrium scenarios extending from interaction quenches\cite{EcksteinKollarWerner_PRL2009, EcksteinKollar_PRL2008}, to DC field-driven systems\cite{thermalizationSciRep2014, NoneqFDT_Frontiers, FreericksPRB2004, FreericksTurkowskiZlatic_PRL2006, Freericks_PRB2008}, to simulations of time-resolved spectroscopies\cite{EcksteinWerner_PRB2011, MoritzEtAl_PRL2013}. However, the effect of disorder in the thermalization of these nonequilibrium systems remains generally understudied.

In this paper we use the recently developed nonequilibrium DMFT+CPA embedding scheme~\cite{NEDMFTCPA_PRB2022} that combines the nonequilibrium extensions of both DMFT\cite{DMFT, DMFT_2, DMFT_3, DMFT_4, DMFT_FK, DMFT_noneq, FK_NonEq_DMFT08, DMFT_noneq_Aoki} and CPA (coherent potential approximation)\cite{CPA_Soven_1967, CPA_Kirkpatrick, CPA_Velicky_1969, CPA_Yonezawa_1973, NonEqCPA_1, NonEqCPA_2}, to investigate the thermalization dynamics of a correlated disordered system modeled by the Anderson-Hubbard model under an interaction quench. In this way, we are able to  assess the impact of the disorder on the relaxation of the system and, specifically, to evaluate the temperature of the system once it has settled into its long-time  thermal state. We analyze the nonequilibrium distribution functions calculated after the quench for various values of the final interaction strengths and as a function of disorder strength. We find that, for moderate interactions after the interaction quench, disorder can tune the final temperature of the system across a broad range of values with increased disorder strength leading to lower effective temperature.




The rest of the paper is structured as follows: In Section \ref{sec:modelmethods}, we briefly discuss the model and review the nonequilibrium DMFT+CPA formalism and its numerical implementation. In Section \ref{sec:results}, we present the results that describe the thermalization of the system after relaxation of the system following the interaction quench. We end the paper with our conclusion in Section \ref{sec:conclusion}.

\section{Model and Methods}
\label{sec:modelmethods}

\subsection{Model}

\noindent We consider a correlated disordered system described by the Anderson-Hubbard model initially in equilibrium at temperature $1/\beta$. The Hamiltonian is given by Eq.(\ref{eq:AndersonHubbard}). Where $t_{ij} = t_{hop}$ is the hopping amplitude between nearest-neighbor sites (denoted by $\langle ij \rangle$), $U(t)$ is the Coulomb interaction strength, and $V_i$ is the random onsite disorder for site $i$. $c^\dagger_{i \sigma}$ and $c_{i \sigma}$ are respectively the creation and the annihilation operators for a particle of spin $\sigma =\uparrow,\downarrow$ at site $i$. $n_{i \sigma}$ is the number of particles of spin $\sigma =\uparrow,\downarrow$ at site $i$ and $\mu$ is the chemical potential. We study the system at half-filling, such that $\mu = U/2$.

\begin{eqnarray}
    \label{eq:AndersonHubbard}
         H  = - \sum_{\langle i j \rangle \sigma} &t_{ij}& (c^\dagger_{i \sigma} c_{j \sigma} + h.c.) + \sum_{i} U(t) n_{i \uparrow} n_{i \downarrow} \nonumber \\
        & + &  \sum_{i \sigma}\left(V_i - \mu \right) n_{i \sigma},
\end{eqnarray}

In equilibrium, the Coulomb interaction is constant $U(t) = U$. In the nonequilibrium scenario of interest in this work,  it is given by a step function $U(t) = \Theta(t-t_{quench}) U_2$ with $t_{quench}=0$, such that the interaction is $U_1 = 0$ for negative times and some constant $U_2 \neq 0$ for positive times. The onsite disorder $V_i$ is constant in time and follows a uniform distribution such that $P(V_i)=\frac{1}{2W}\Theta(W-|V_i|)$, where $W$ is the disorder strength. We use the notation $\langle ... \rangle_{\{V\}}$ to indicate averaging over all disorder values in the angle brackets. Here, we focus on the model for the Bethe lattice in the limit of infinite coordination number.

\subsection{Nonequilibrium DMFT+CPA}
\noindent The nonequilibrium many-body formalism can be formulated on the Keldysh contour whereby the system is evolved forward in time from an early $t = t_{min}$ to times of physical interest up to a maximum value $t_{max}$
and then back backward to the early times again\cite{Keldysh64_65, StefanucciLeeuwen_CUP2013, rammer_2007}. The formalism involves several types of two-time Green's functions among which  $G^<(t,t')$  (the lesser),  $G^>(t,t')$ (the greater), and $G^R(t,t')$ (the retarded) Green's functions. In the context of a system initially in equilibrium at an initial temperature $T=1/\beta$, a vertical spur of imaginary times of length $-i\beta$ is added to the Keldysh contour resulting in the so called Kadanoff-Baym-Keldysh contour\cite{Keldysh64_65, BaymKadanoff62}. In this situation, one should add to the previous types of Green's functions in the formalism, the Matsubara Green's function $G^{\tau}$, and the mixed time Green's functions, where one of the times is on either one of the horizontal branches of real times, while the other is on the vertical branch of imaginary times. The solution for a given problem can be either formulated in terms of the different Green's functions $G^<$, $G^>$ $G^R$, $G^{\tau}$, etc. Alternatively, it can be formulated in terms of the contour-ordered Green's function $G_c(t,t')$ from which all the others can be  extracted. It is this latter approach that we use in this work. The contour-ordered quantities have time ordering performed with respect to time advance along the entire contour. Hereafter we drop the subscript $c$ from the contour-ordered quantities for convenience.


Our solution for the above described Anderson-Hubbard model under an interaction quench is performed within the recently developed nonequilibrium DMFT+CPA formalism which builds on the equilibrium formalism\cite{DMFT_CPA_1, DMFT_CPA_2, DMFT_CPA_3, DMFT_CPA_4, DMFT_CPA_5} and is described extensively in Ref.[\onlinecite{NEDMFTCPA_PRB2022}]. Here, for the sake of completeness, we briefly summarise the algorithm. The method maps the lattice problem onto that of an impurity embedded in a self-consistently determined medium characterized by the hybridization $\Delta(t,t')$ that is consistent with that of DMFT for the clean system and with that of the disordered non-interacting system for CPA.

In practice, the algorithm consists of the following self-consistency procedure. From an initial guess of the hybridization function $\Delta(t,t')$, one obtains the noninteracting Green's function for each disorder configuration given by:
\begin{equation}
    \mathcal{G}_{V_i}(t,t') = \left( \left(i\partial_t + \mu -V_i \right)\delta_c - \Delta) \right)^{-1}(t,t')
\end{equation}
\noindent From this, one obtains the Coulomb interaction self-energy. Here, similar to Ref[\onlinecite{NEDMFTCPA_PRB2022}], we focus on the weak-to-moderate interaction and disorder strengths regime, and we use second order perturbation theory so that the self-energy is given by:
\begin{equation}
\label{eq:sigmaU} 
 \Sigma_{V_i}(t,t') = -U(t)U(t') \mathcal{G}_{V_i}(t,t')^2 \mathcal{G}_{V_i}(t',t).
\end{equation}
After obtaining the self-energy for all disorder configurations, we evaluate the disorder-averaged Green's function:
\begin{equation}
 G_{ave}(t,t')  =  \langle \left(G_{V_i}\right) \rangle_{\{V\}}
 \end{equation}
 where $G_{V_i}(t,t')$ is the Green's function for the disorder configuration $\{V_i\}$:
 \begin{equation}
 G_{V_i}(t,t')  =  \left[ \mathcal{G}_{V_i}^{-1} - \Sigma_{V_i} \right]^{-1}(t,t').
\end{equation}
This is followed by the evaluation of the updated hybridization function which in the present case of the Bethe lattice with infinite coordination is given by $\Delta(t,t') = {t^*}^2 G_{ave}(t,t')$ and the self-consistency loop is repeated starting from the calculation of the new Coulomb interaction self-energies and proceeds until convergence of the self-energy within a desired criterion. $t^*$ is the hopping amplitude rescaled with the coordination number $z$ so that $t_{hop} = \frac{t^*}{\sqrt{z}}$. We use $t^*=0.25$ and thus set the bandwidth to be our energy unit and its inverse to be the time unit. 






\begin{figure}[htbp]
\includegraphics[scale=0.37]{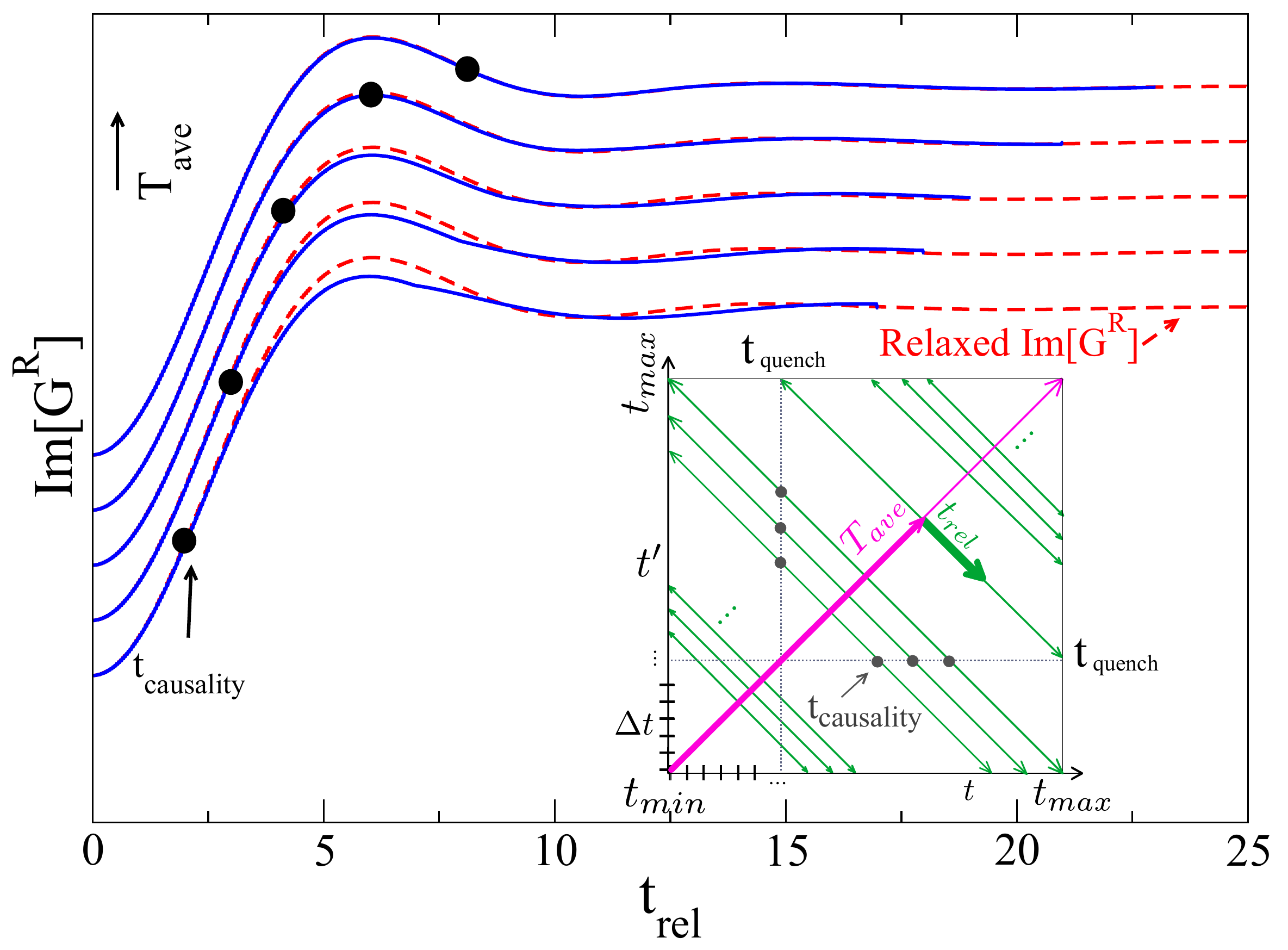}
\caption{Imaginary part of the retarded Green's function as a function of relative time for a range of average times, for $U_2 = 3t^*, \; W = t^*$. The long-time (relaxed) retarded Green's function is represented by the dashed red lines, and the causality time is marked for each average time with a black dot. Note that the $t_{rel}$ at which $G^R$ begins to diverge from its relaxed form is greater and greater with increasing $T_{ave}$. Inset: illustration of the relationship between $(T_{ave}$, $t_{rel})$ and $(t, t')$, with the blue vertical and horizontal lines indicating the time at which the quench occurs in $t$ and $t'$ and black dots indicating $t_{causality}$.}
\label{fig:GR_Trel}
\end{figure}

\subsection{Numerical Implementation}

\noindent Our implementation of the nonequilibrium DMFT+CPA follows the discrete time construction of Refs.[\onlinecite{NEDMFTCPA_PRB2022, Freericks_PRB2008}]. The Kadanoff-Baym-Keldysh contour is discretized into $(2N_t + N_{\tau})$ time steps, where $N_t$ is the number of time steps on each leg of the horizontal real-time branch of the contour and $N_{\tau}$ is the number of time steps on the vertical branch of imaginary time. The step sizes are  $\Delta t = \left(t_{max} - t_{min}\right)/N_t$ for real time and $\Delta \tau = \beta/N_{\tau}$ for imaginary time. In this paper, $t_{min}=-5$ and $t_{max}=20$ while the initial temperature of the system is such that $\beta_{initial}= 15$.

In this context, the contour-ordered quantities such as $G(t, t')$ become square complex matrices $G_{ij}$ of size $(2N_t + N_{\tau}) \times (2N_t + N_{\tau})$. Convolutions of contour-ordered quantities becomes matrix multiplications, and the continuous matrix inverse becomes a discrete matrix inverse.
The analysis is often performed by switching from the $(t,t')$ time coordinates to the Wigner coordinates $(T_{ave}, t_{rel})$ where $T_{ave}$ can be viewed as the effective time of the system while frequency domain information is obtained by Fourier transforming with respect to $t_{rel}$. Observables calculated from the discretized contour, such as the distribution functions and the energy, are often obtained for multiple step sizes then extrapolated to the continuum limit $\Delta t \to 0$. We use standard Lagrange interpolating polynomials to quadratic order.


\begin{figure}[t] 
\includegraphics[scale=0.35]{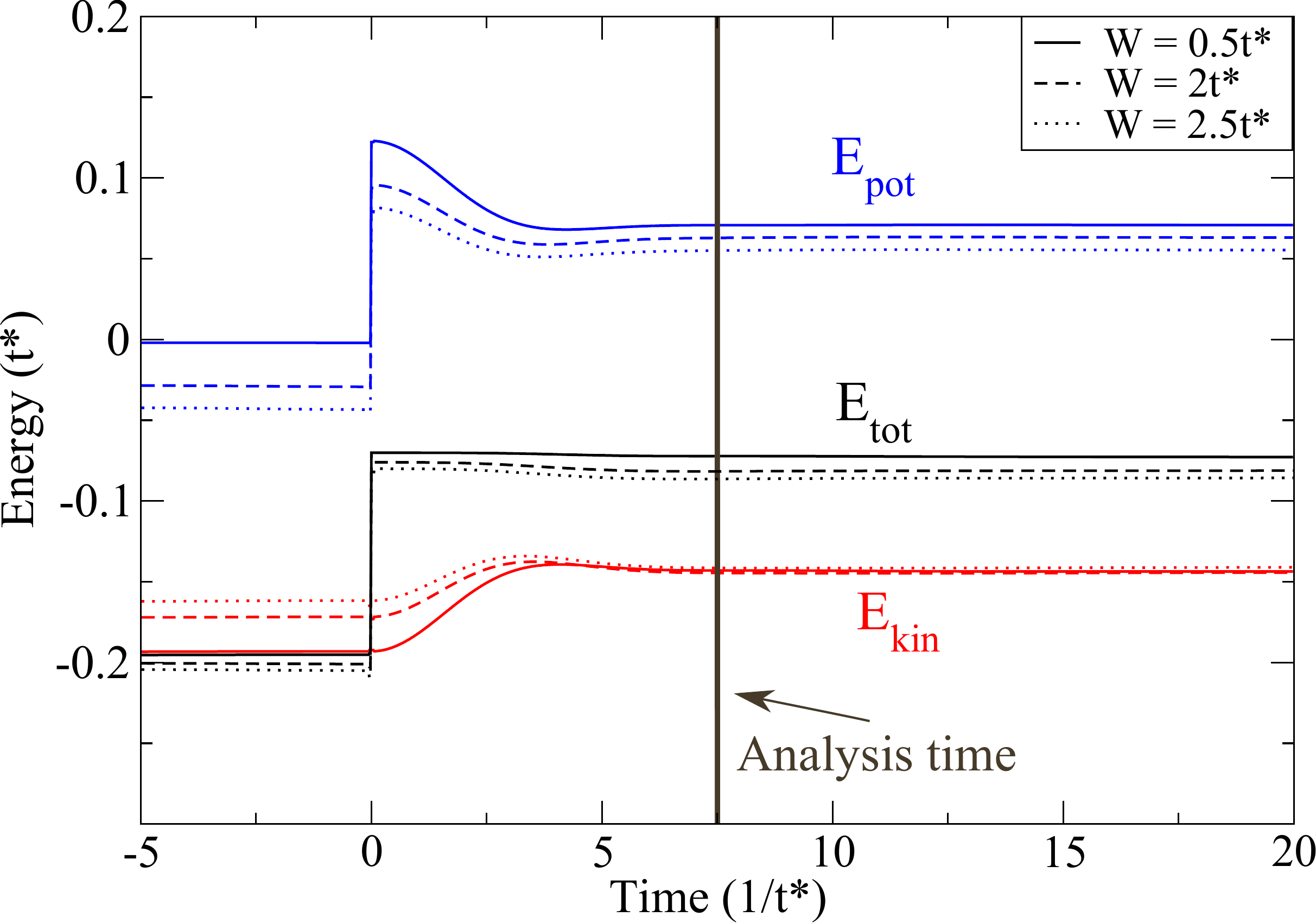}
\caption{Extrapolated potential, kinetic, and total energies for $U_2 = 2t^*$. The vertical black line shows the time at which we evaluate the relaxed distribution function. This time is well after the relaxation of the system.}
\label{fig:energies}
\end{figure}

\section{Results}
\label{sec:results}
\noindent The system is initially in equilibrium at temperature $T=1/\beta$ with $\beta_{initial} = 15$. While keeping the disorder strength $W$ constant, the interaction quench is applied at time $t=t_{quench} = 0$ with the interaction abruptly changing from an initial value $U_1=0$ to a final value $U=U_2$. We are interested in tracking the thermalization of the system at long times. Our analysis is guided by two fundamental quantities: the density of states and the distribution function. For a thermalized system, the former is given by the retarded Green's function, while the latter is given by the lesser Green's function. Namely:
\begin{equation}
    \rho(\omega) = -i \mathrm{Im} G^R(\omega)/\pi
\end{equation}
and according to the fluctuation dissipation theorem, for a thermalized system,
\begin{equation}
    G^<(\omega) = -2 i F(\omega) \mathrm{Im} G^R(\omega).
    \label{eq:FDT}
\end{equation}
Where $F(\omega)$ is the distribution function. In the nonequilibrium formalism, we can track these quantities as a function of average time.

\subsection{Density of states}
\noindent For the system at half-filling, we know that the real part, in the time domain, of the retarded Green's function vanishes for all average times~\cite{thermalizationSciRep2014}. Thus, the density of states is fully defined by the imaginary part of the retarded Green's function in the time domain. For this reason, we can track the dynamics of the density of states through the imaginary part of the retarded Green's function in the time domain. Fig.~\ref{fig:GR_Trel} presents the typical behavior of $G^R(T_{ave}, t_{rel}) $ as a function of $t_{rel}$ for a series of $T_{ave}$ values. Note that the relative time axis (represented by the green lines in the insert), for earlier values $T_{ave}$ (magenta lines in the insert), has segments of time coordinates $(t,t')$ for which one (or both) of the times is (are) before the interaction quench leading to a mixed character of the corresponding $t_{rel}$ coordinates. The blue lines in the main figure correspond to successive $T_{ave}$ values after the quench, while the dashed red line corresponds to an average time value after the quench for which all $t_{rel}$ involves both $t$ and $t'$ that have the new interaction strength $U_2$. The black circles correspond to the causality time beyond which $t_{rel}$ has mixed character. One can see on this figure that the solid blue curves overlap with the dashed red curve up to the causality time and that the retarded Green's function is only constrained by causality. So, the density of states of states is immediately established after the quench. The relaxation of the system can thus be tracked through the distribution function.

\begin{figure}[t] 
\includegraphics[scale=0.35]{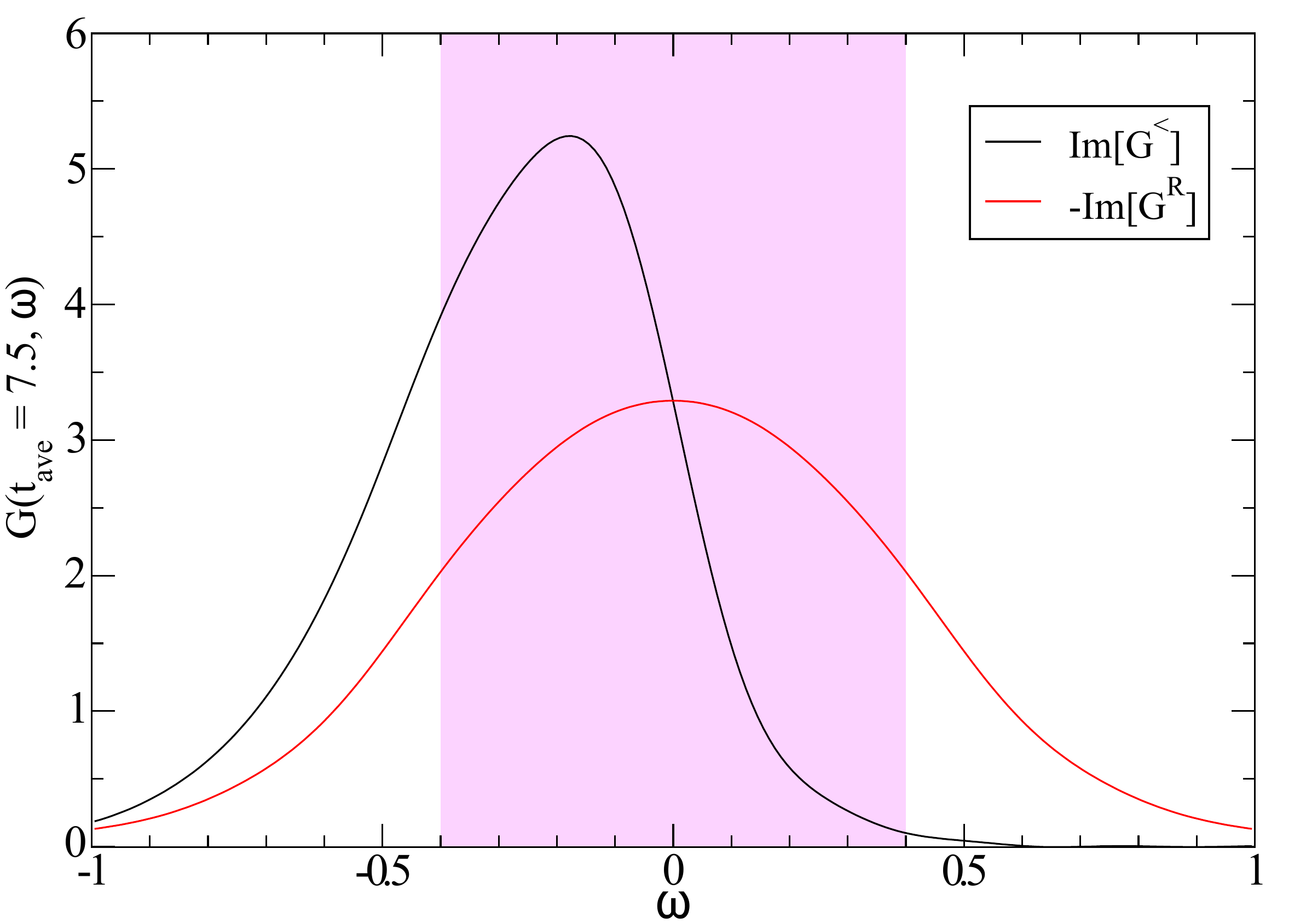}
\caption{Imaginary parts of lesser and retarded Green's functions as a function of frequency for $U_2 = 2t^*,\; W = 2t^*$ at the analysis time. The shaded box shows the region over which we evaluate the distribution function $F(\omega) = -Im[G^<]/(2 Im [G^R])$. Outside of this region, the ratio is prone to numerical instabilities due the Gibbs phenomenon in the frequency data obtained and to the division by small numbers.}
\label{fig:GLGRbox}
\end{figure}

\subsection{Distribution function $F(\omega)$}
\noindent In the present study, we are interested in the thermalization of the system after it has undergone its early transient following the quench. Fig.~\ref{fig:energies} shows, for different disorder strengths and for $U_2 = 2t^*$, the evolution in time of the kinetic, potential and total energies of the system evaluated following Refs.[ \onlinecite{NEDMFTCPA_PRB2022, HubbardQuenchEckstein}]. The quench is performed at time $t=0$. After an initial nontrivial response to the quench, the observables settle into a constant value for the remaining duration of the simulation. The vertical black line indicates the time $t=7.5$ at which the long-time analysis is performed.

Given that the density of states is established immediately after the quench and is only constrained by causality, this analysis time is chosen so as to allow a range of $t_{rel}$ values that enables a reliable Fourier transform. To obtain the distribution function, we will use the fluctuation-dissipation theorem as expressed by Eq.(\ref{eq:FDT}). To this end, we first Fourier transform the lesser and retarded Green's functions $G^{R/<}(T_{ave}, t_{rel})$ in relative time to yield $G^{R/<}(T_{ave}, \omega)$. The result of this operation is illustrated for $U_2 = 2t^*$ and $W=2t^*$ in Fig.~\ref{fig:GLGRbox}. To avoid numerical instabilities, the distribution function is only evaluated in a frequency range around $\omega = 0$ for which both $G^R(T_{ave}, \omega)$ and $G^<(T_{ave}, \omega)$ remain finite as illustrated by the shaded box in Fig.~\ref{fig:GLGRbox}. 






\begin{figure}[t] 
\label{fig:nonthermalfw}
\includegraphics[scale=0.345]{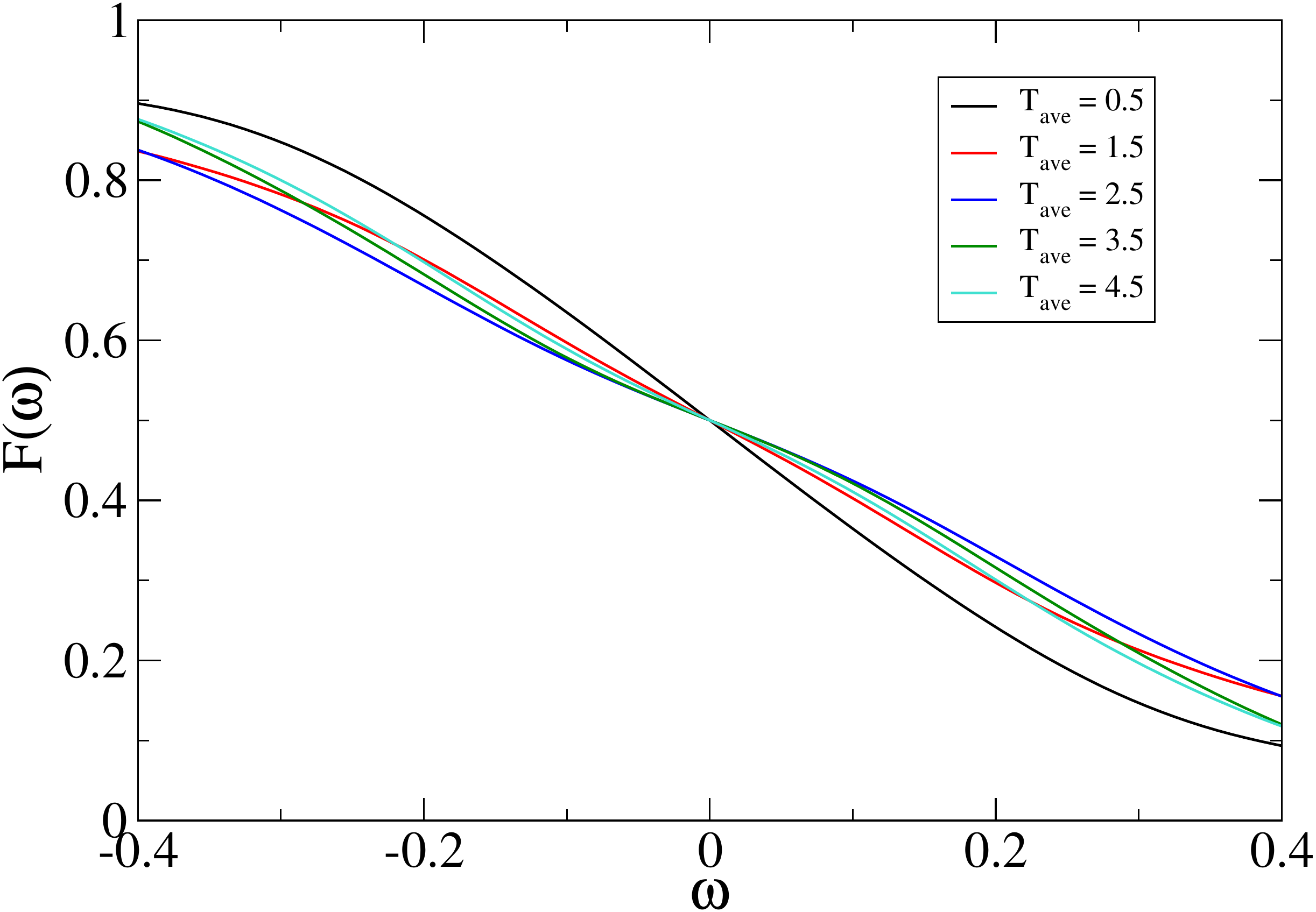}
\caption{Relaxation of $F(\omega)$ at $U_2 = 3t^*, W = t^*$ soon after the quench, but before thermalization, demonstrating the non-thermal form of the distribution function at the early stages of the relaxation.}
\label{fig:distrFunction1}
\end{figure}

\begin{figure}[h]
\includegraphics[scale=0.345]{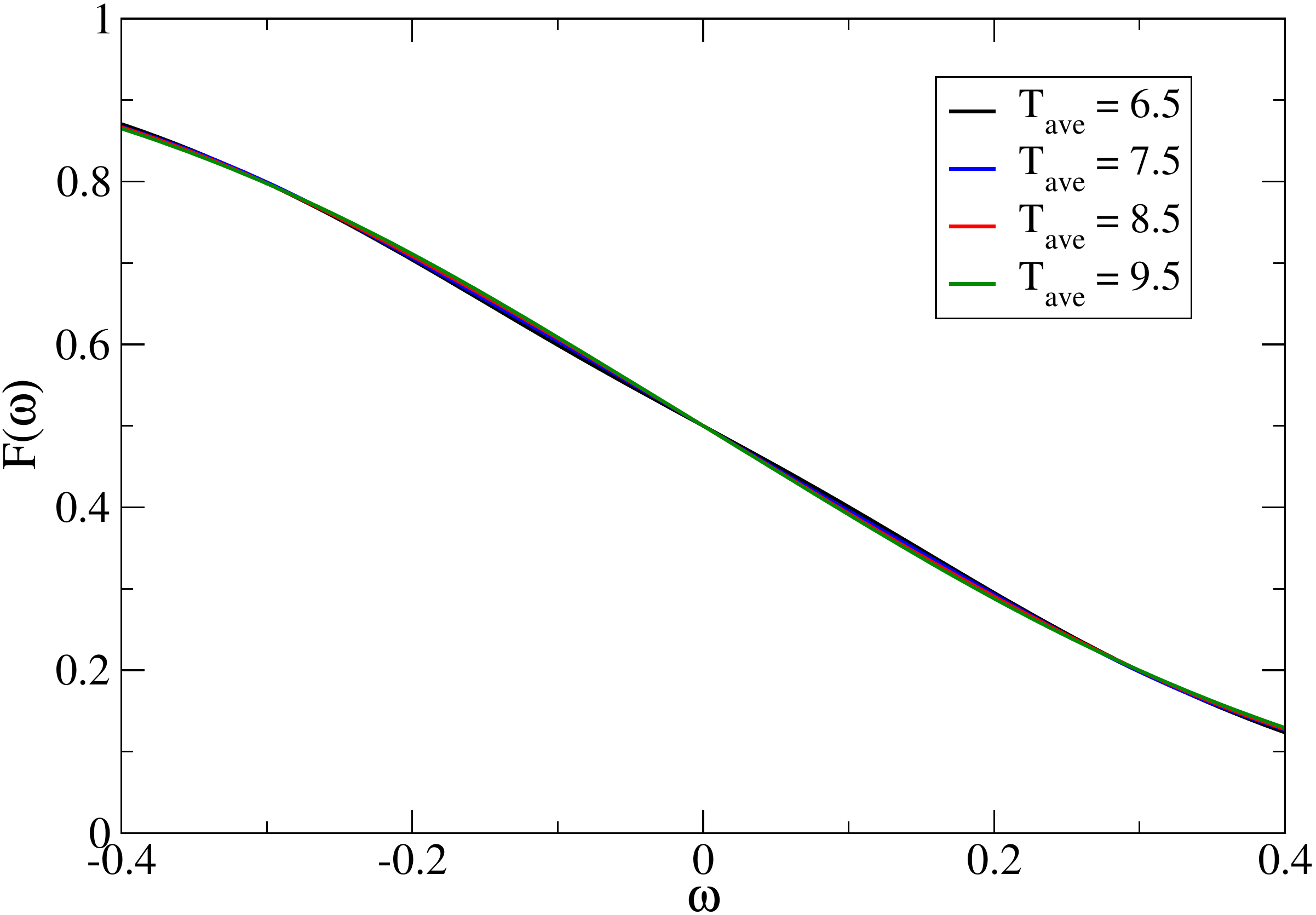}
\caption{Post-relaxation $F(\omega)$ for times slightly before and slightly after our analysis time for $U_2 = 3t^*, W = t^*$, demonstrating that the distribution function changes minimally around this analysis time at which we evaluate the relaxed $F(\omega)$.}
\label{fig:distrFunction2}
\end{figure}

Figs.~\ref{fig:distrFunction1} and (\ref{fig:distrFunction2}) show the extracted distribution function for $U_2 = 3t^*$ and $W=t^*$ for different average times. One can readily observe that following the interaction quench at time $t=0$, the distribution function initially changes in a highly nontrivial way and may in fact clearly correspond to a non-thermal system (Fig.~\ref{fig:distrFunction1}). However, around our analysis time, corresponding to $T_{ave}=7.5$, the distribution function is seen to change very little for different values of the average time and the different curves essentially overlap (Fig.~\ref{fig:distrFunction2}). For this reason, the system can be assumed to have settled into its long-time state at time $T_{ave}=7.5$. It is in this regime that we evaluate a long time effective temperature of the system after the quench.\\

\begin{figure}[t] 
\includegraphics[scale=0.34]{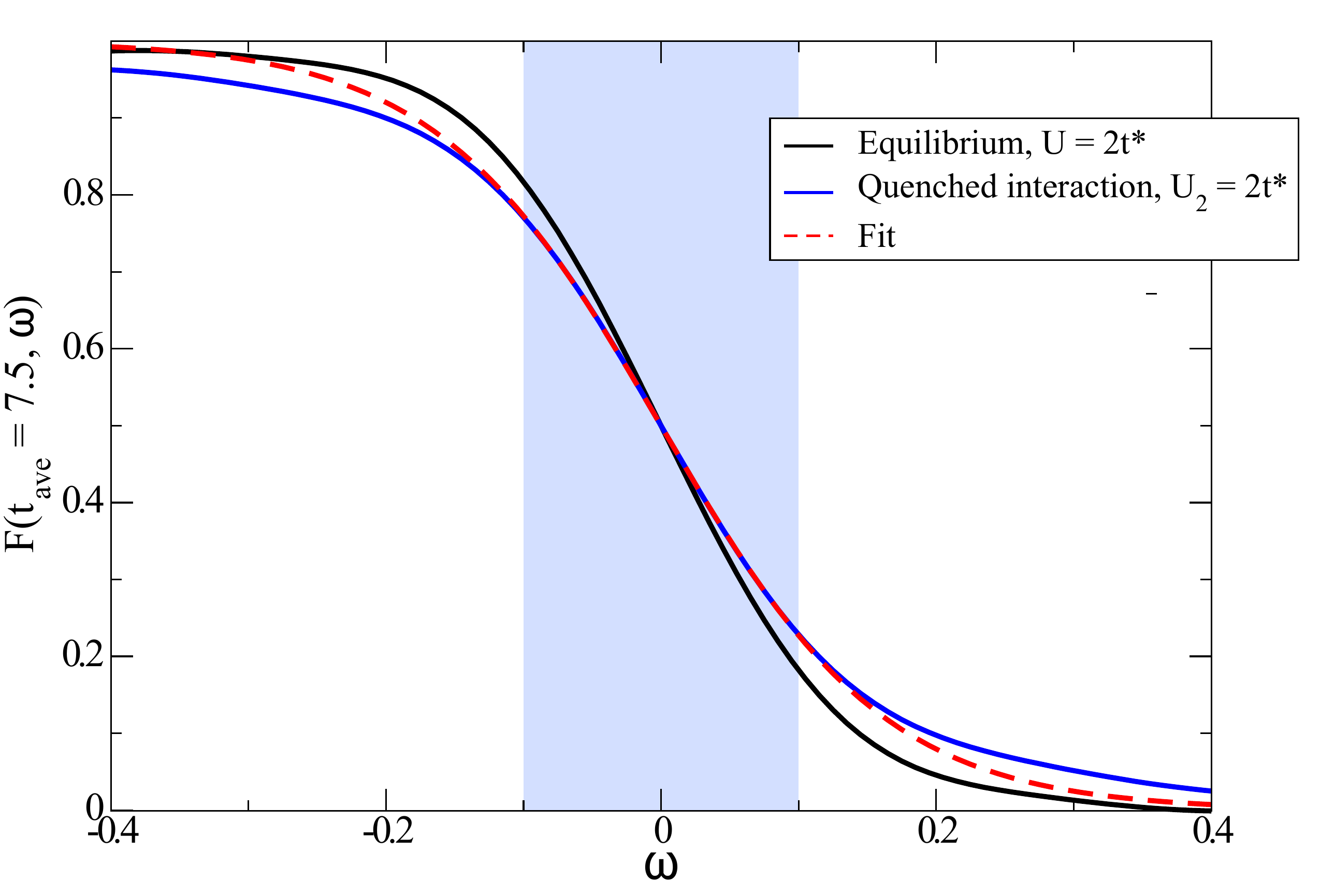}
\caption{Distribution function for the equilibrium system with $U = 2t^*, \; W = 2t^*, \; \beta = 15$, and after relaxation for the quenched system with $U_2 = 2t^*, \; W = 2t^*, \; \beta_{initial} = 15$. The dashed line shows the fit to the quenched system distribution function after the transient. The shaded box indicates the region over which the fit is performed. Here we fit the Fermi function, $F_{Fit}(\omega)= 1/(1+\mathrm{exp}\left(\beta \omega ) \right)$ with $\beta$ as a free parameter, to the calculated $F(\omega)$, and this allows us to extract an effective temperature.}
\label{fig:fwquencheqfit}
\end{figure}

\begin{figure}[t] 
\includegraphics[scale=0.34]{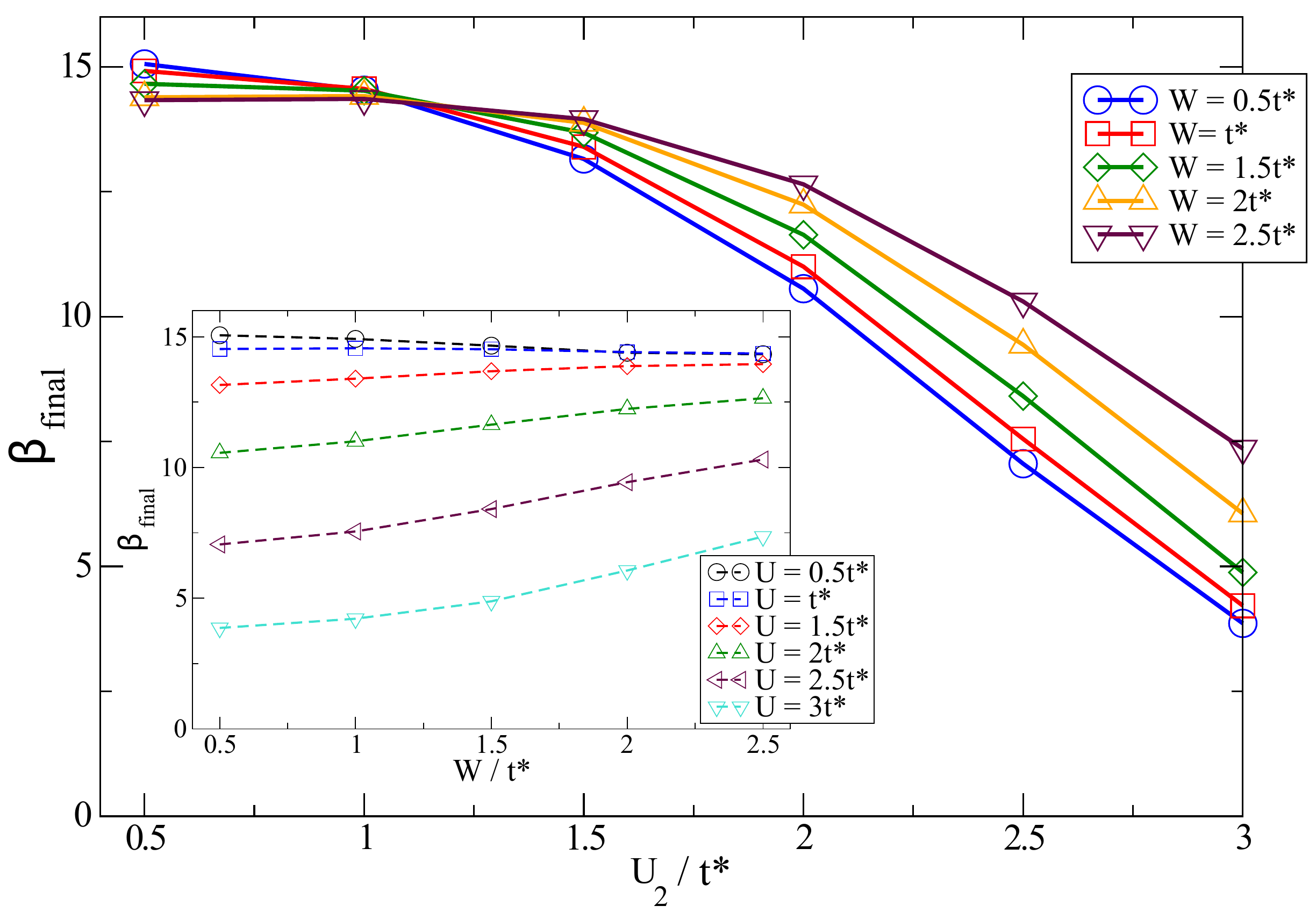}
\caption{Inverse effective temperature as a function of the final interaction strength for different disorder strengths. The systems is initially at a temperature such that $\beta_{initial} = 15$. Inset: Effective inverse temperature $\beta$ as a function of the disorder strength for different interaction strengths. Increased disorder strength for moderate interaction strengths leads to a lower long-time temperature.}
\label{fig:betavsWandU}
\end{figure}


\subsection{Effective temperature}
\noindent The effective temperature is obtained by fitting a Fermi-Dirac distribution function ($F(\omega)= 1/(1+\mathrm{exp}\left(\beta \omega)\right)$ with $\beta$ as a free parameter) to the extracted distribution function over a frequency window around $\omega = 0$ as illustrated in Fig.~\ref{fig:fwquencheqfit}. 
As indicated above, after the quench but before relaxation, the distribution function can take non-thermal forms ~\textcolor{red}{(}Fig.\ref{fig:distrFunction1}). Consequently, an effective temperature cannot be traced over the entire time evolution of the system. However, this procedure is well-defined for the chosen analysis time for the long-time behavior.

Fig.~\ref{fig:betavsWandU} shows the long time effective temperature of the system as a function of the final interaction strength $U_2$ with different solid lines corresponding to different values of the disorder strength $W$. The inset shows the same data but with the disorder strength on the $x$-axis and different dashed lines corresponding to different values of the final interaction strength. 
The figures show the significant dependence of the final inverse temperature on disorder strength. For weak $U_2$ values, increased disorder strength leads to small increase in the long time temperature. However, as the interaction strength $U_2$ is increased, we observe that increasing the disorder strength leads lower long-time effective temperatures. This shows that under an interaction quench, the long-time temperature can vary over a broad range of values depending on the disorder strength, with increased disorder strength leading to lower final temperature.

\section{Conclusion}
\label{sec:conclusion}

\noindent We have analyzed the relaxation of a disordered interacting system after an interaction quench where, with the disorder strength held constant, the interaction strength is abruptly switched from zero to a finite value $U_2$ at which it is subsequently kept. We have used the recently developed nonequilibrium DMFT+CPA formalism that maps the lattice problem onto an effective mean field that is equivalent to that of the dynamical mean field theory (DMFT) for the clean system and to that of the coherent potential approximation (CPA) for the disordered noninteracting system. By extracting the distribution function from the Green's function using the fluctuation-dissipation theorem, we showed that while the early transient does not follow the fluctuation dissipation theorem, at longer times, the system settles into a thermal state at a constant temperature. This long time temperature is lowered by increased disorder strengths at moderate values of the interaction.
Altogether our studies demonstrate that after the interaction quench, disorder can tune the long-time temperature of the system over a broad range of values.

\section*{Acknowledgments} 
\noindent HFF is supported by the National Science Foundation under Grant No. PHY-2014023. HT has been supported by NSF DMR-1944974 grant.


\end{document}